\documentclass[runningheads]{llncs}
\usepackage[T1]{fontenc}
\usepackage{tcolorbox}
\usepackage{tabularx}
\usepackage{amsmath}
\usepackage{multirow}
\newcolumntype{Y}{>{\centering\arraybackslash}X}
\usepackage{graphicx}
\begin{document}
%
\title{Automated Grading of Handwritten Mathematics Using Vision-Capable LLMs}
\titlerunning{Auto grading of Handwritten Math Using Vision-Capable LLMs}
%
\author{Jacob Levine\inst{1}\orcidID{0009-0003-2530-7690} \and
Miguel Aenlle\inst{1}\orcidID{0009-0002-3082-4441} \and
Craig Zilles\inst{1}\orcidID{0000-0003-4601-4398} \and
Matthew West\inst{1}\orcidID{0000-0002-7605-0050} \and
Mariana Silva\inst{1}\orcidID{0000-0001-6087-7179}}
\authorrunning{J. Levine et al.}
%

\institute{University of Illinois Urbana-Champaign\\
\email{\{jlevine4,maenlle2,zilles,mwest,mfsilva\}@illinois.edu}
}

\maketitle              
\begin{abstract}
Automated grading systems have enabled scalable assessment for many response types, but handwritten mathematics remains a barrier due to the complexity of multi-step solutions. Vision-capable large language models (LLMs) offer new opportunities here, yet their reliability in authentic instructional settings remains poorly understood.

We present an empirical evaluation of an LLM-based grader for handwritten mathematical work using instructor-defined rubrics. Extending a prior pipeline for typed responses, we integrate transcription and rubric-based evaluation of photographic submissions within a single LLM call, evaluating on student work from two university STEM courses. Comparing AI grading decisions against human-assigned ground truth at the rubric-item level, we observe high overall accuracy, with most errors --- 87\% in the best model --- attributable to transcription failures rather than rubric misapplication. We categorize common error modes, including image quality issues, hallucinated content, and incorrect handling of equivalent expressions. These findings highlight both the promise and limitations of LLM-based grading for handwritten mathematics, providing guidance for system design, prompt refinement, and deployment in educational settings.

\end{abstract}
\section{Introduction and Background}

Online question-asking systems offer benefits including reduced grader workload, immediate feedback, enhanced practice through question variants, and support for secure asynchronous exams with minimal cheating risk due to randomization \cite{west_integrating_2021,silva_cheating_2020}. However, many instructors avoid these systems due to limited input options. While most platforms support multiple choice, numerical answers, and code autograding, they typically require a single ``correct'' answer or deterministic grading function, making proof-based questions nearly impossible with rule-based approaches \cite{poulsen_proof_2022}.

Large Language Models (LLMs) have enabled automated graders capable of more nuanced judgments. In practice, LLMs can often evaluate logical steps or whether statements fulfill specific properties, allowing them to identify errors in student logic or grade against rubrics \cite{zhao_sigcse2026_toappear}. However, LLM-based approaches remain insufficient for mathematics courses where students must hand-write solutions. Accordingly, we evaluate vision-capable LLMs for rubric-based grading of handwritten mathematical work from authentic university courses.


Proposals for LLM-based graders for text answers are fairly well-studied. Zhao et al. \cite{cristea_language_2025} examined GPT-4o and GPT-o1-preview across different prompting strategies, finding that while directly asking an LLM to grade submissions based on internal knowledge is generally unsuccessful, clever prompting can yield performance comparable to human graders. Specifically, on a dataset of induction proofs, LLMs achieved strong results when grading against a rubric and when augmented with retrieval of similar graded examples. Similarly, Li et al. \cite{li_llm-based_2025} designed a human-in-the-loop system where the LLM receives questions and a rubric, then iteratively refines the rubric by querying human graders, achieving state-of-the-art performance.

There has been extensive work building AI models to recognize mathematical expressions, including handwritten mathematical expression recognition (HMER) \cite{truong_HMER-survey_2024,xie_icdar-CROHME_2023}.  In general, recognition is improved when stroke data is available (e.g., when writing is captured on a digital tablet) than images~\cite{truong_HMER-survey_2024}.  Recent work found that purpose-built systems (e.g., Donut-style~\cite{kim_donut_2022} visual-encoder text-decoder systems) had higher recognition accuracy than frontier multi-modal LLMs (i.e., GPT-4o) at the time~\cite{wang_image-over-text_2025}, but those results pre-date the most recent LLM releases (e.g., Gemini 3).

In addition, researchers have attempted grading images of handwritten math directly. Liu et al. scanned traditional handwritten paper exams, translating handwritten math to LaTeX using GPT-4V and the proprietary software MathPix, then checking answers against rubric rules using GPT-4, achieving grading accuracy of $\sim60\%$~\cite{liu_ai-assisted_2024}.  They find that reading student work (recognition and layout) is a primary challenge in LLM grading~\cite{kortemeyer_handwritten-thermo_2024}.  
Kortmeyer et al. graded paper exam scans using images of the rubrics given to TAs using GPT-5 and found the accuracy of unfiltered AI to be only adequate for low-stakes feedback ($R^2 = 0.85$ with TA grades)~\cite{kortemeyer_handwritten-calculus_2025}.  Their ``human-in-the-loop'' grader using an IRT-based confidence method was able to achieve higher agreement ($R^2 = 0.85$), but at the cost of manual review of 70\% of items.


Our work differs from prior work in several key ways. First, students are responsible for capturing images of their handwritten solutions and can review and resubmit if the image quality is insufficient, which may contribute to the higher accuracy we observe. Second, rather than treating grading errors as a single category, we explicitly distinguish between transcription errors and rubric-application errors. Third, our approach is fully automated and supports randomized problem instances across students.


Building on these design choices, this paper makes the following contributions: (1) extending an existing rubric-based LLM grading pipeline to support photographic submissions of handwritten mathematics using a single multimodal LLM call for both transcription and rubric evaluation; (2) evaluating three vision-capable LLMs on authentic student submissions from two STEM courses across five open-ended problems; and (3) presenting a rubric-item-level error analysis that distinguishes transcription failures from rubric-application failures and identifies practical mitigation strategies.




\section{Methods}

We developed and deployed a prototype AI-assisted grading tool integrated into PrairieLearn\footnote{\url{https://www.prairielearn.com}}, an open-source online assessment platform. The tool evaluates open-ended mathematical responses submitted as handwritten work and supports both manual and automated grading workflows. Although the platform allows multiple submission attempts, only the final uploaded image is graded. Our grader was originally designed for typed responses \cite{zhao_sigcse2026_toappear}; in this work, we extend that pipeline to support handwritten submissions by introducing an image-based grading pipeline and prompt structure that jointly perform transcription and rubric-based evaluation.

In the modified pipeline, the LLM prompt\footnote{Code available at \url{https://github.com/PrairieLearn/PrairieLearn}} includes the problem statement, rubric, optional reference solutions and instructions, and explicit directives for both transcription and assessment. This single-call design is motivated by prior findings showing higher accuracy compared to separating transcription and grading \cite{levine_toappear}. The AI grader operates asynchronously, executing only after submissions close and the grading configuration is finalized. Human graders may review and override AI evaluations, and human-assigned grades always take precedence.


To evaluate LLM performance on grading handwritten mathematics, we conducted a study using student solutions from two courses at a large Midwestern research university: a numerical methods course (Course 1, $n=440$) and an introductory dynamics course (Course 2, $n=245$). Both courses required mathematical derivations submitted as photographs but differed in submission methods, grading policies, and proctoring conditions. We tested three questions from Course 1 and two from Course 2. 

\begin{figure}[h]  
    \centering  
    \includegraphics[width=1\linewidth]{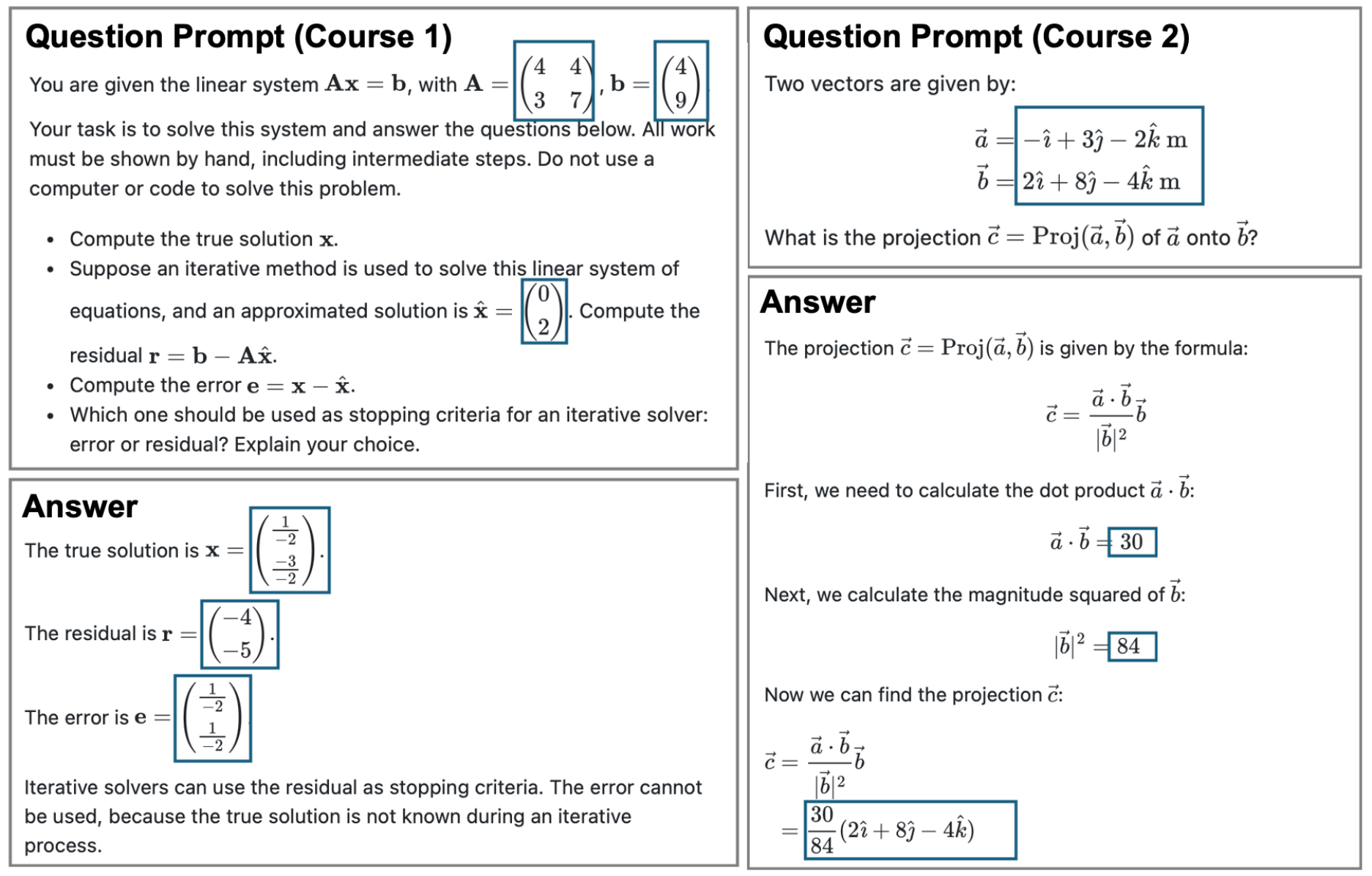}  
    \vspace{-8mm}
    \caption{Example question HTML and reference answers in the LLM prompt, with randomized parameters and corresponding correct values highlighted.} 
    \vspace{-4mm}
    \label{fig:examplequestion}  
\end{figure}

Questions were templated and randomized using PrairieLearn, affecting both the problem text and reference answers (Figure~\ref{fig:examplequestion}). While this randomization reduces opportunities for academic misconduct \cite{silva_cheating_2020}, it precludes retrieval-augmented grading, as the variability across problem instances prevents reliable reuse of previously graded examples.
 
In Course 1, the questions were part of an extra-credit, in-lecture activity that students completed in groups of three. Students submitted their work as images captured using either a laptop webcam or a mobile phone. The questions spanned distinct concepts and required varied answer formats: error versus residual in an iterative solver (Fig.~\ref{fig:examplequestion}(left)), steady-state computation in a Markov process, and Hessian matrix calculation. In Course 2, both questions appeared on proctored midterm exams. Students wrote their answers on scratch paper and uploaded using attached document cameras. In the first question (Fig.~\ref{fig:examplequestion}(right)), students were asked to compute the projection of one vector onto another. In the second question, students analyzed the planar motion of a rigid body.

\begin{figure}[h]
    \centering
    \includegraphics[width=1\linewidth]{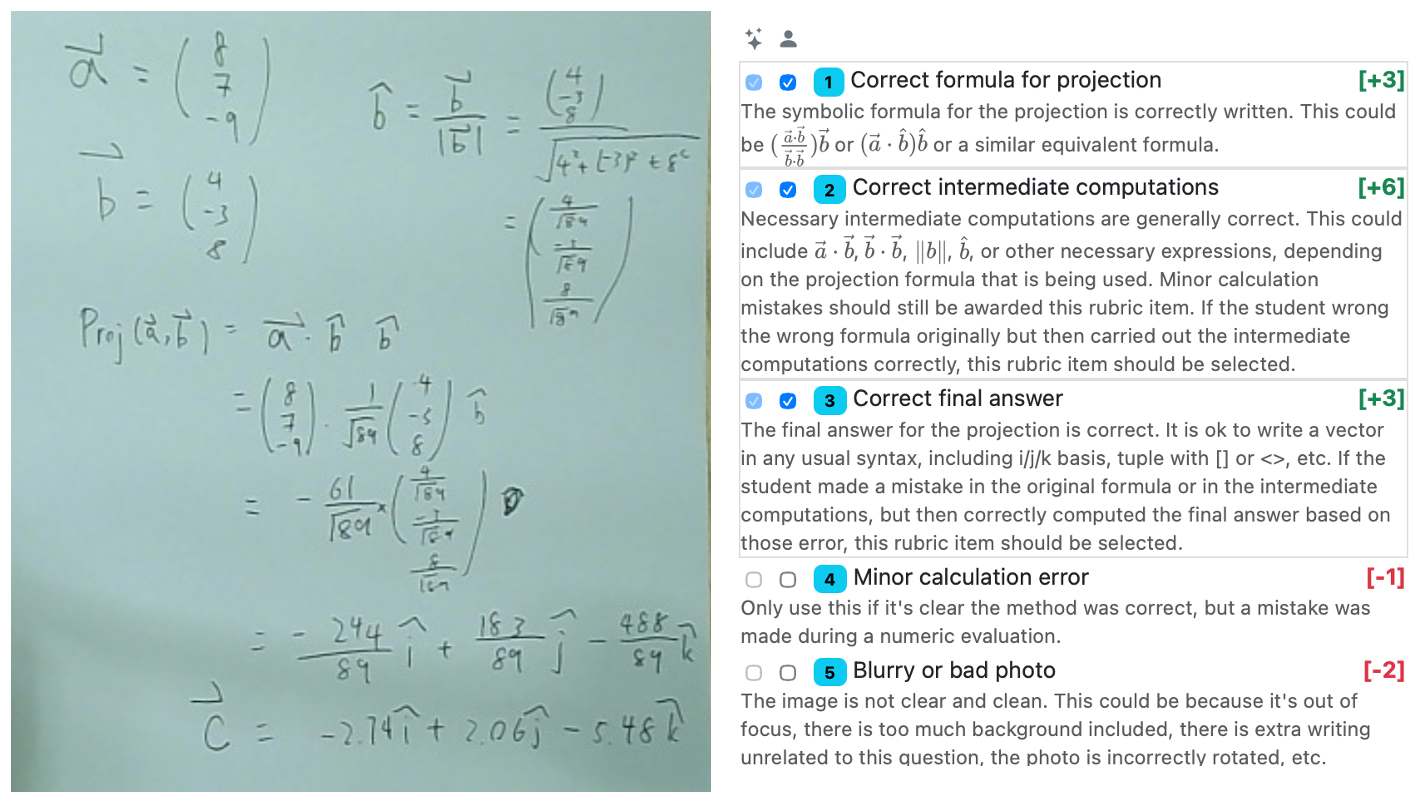}
    \vspace{-7mm}
    \caption{Example of student submission and rubric for Course 2}
    \vspace{-4mm}    
    \label{fig:subs2}
\end{figure}

In both courses, students were able to preview their submission to ensure readability and to crop the image to include only the work relevant to the question. Each submission was graded using a rubric designed by the instructor for the corresponding problem, which evaluated both the correctness of the final answer and the intermediate steps used in the solution. Examples of student submissions and rubric are illustrated in Figure~\ref{fig:subs2}.

Course instructors (who were also authors of this study) graded all submissions for each question, both to assign student grades and to establish a ground-truth reference for this study. Students were not exposed to any AI-generated grading output, which was used solely for the purposes of this research.

To assess the performance of the grading pipeline for each LLM model, we compared the AI-generated grading decisions to the ground-truth human grades at the level of individual rubric items. 
\vspace{-3mm}
\begin{itemize}
    \item \textbf{Rubric-Item Accuracy (RIA):} The proportion of rubric items for which the AI grader’s evaluation matched the human-assigned grade.  
    \item \textbf{False Positives (FP):} 
    The proportion of rubric items that the \textbf{AI selected} but the \textbf{human grader did not select}.
    \item \textbf{False Negatives (FN):} The proportion of rubric items that the \textbf{AI did not select} but the \textbf{human grader selected}.
\end{itemize}

For each disagreement (FP and FN), we examined the model’s output text to determine whether the error was attributable to transcription failures or to incorrect application of rubric logic:

\begin{itemize}
    \item \textbf{Transcription Error (TE):} The proportion of rubric items with AI–human disagreement associated with transcription errors, including missing text that would satisfy a rubric item, hallucinated text, or mistranscribed expressions that render a correct solution incorrect.
    \item \textbf{Rubric Application Error (RAE):} The proportion of rubric items with AI–human disagreement associated with errors in rubric application or justification, such as rubric rules not being followed.
\end{itemize}

In both courses, we performed a complete categorization of each rubric item with AI–human disagreement. Additionally, we conducted qualitative analysis to identify the likely causes of observed errors and to assess whether these errors could plausibly be mitigated through changes to the prompt, rubric structure, or grading pipeline. 

We tested three LLM models: OpenAI's GPT-5-mini and GPT-5.1 and Google's Gemini-3-flash. All of these models have demonstrated relatively advanced vision capabilities, have been tuned to perform step-by-step explanations \cite{singh_GPT5_2025}, and have costs-per-token that allow these models to feasibly be used in a production autograder.

\section{Results and Discussion}

\begin{table}[!htb]
\centering
\caption{Rubric-item accuracy and error breakdown by question and LLM model. Course 1 is labeled as C1 and Course 2 is labeled as C2.}
\vspace{-2mm}
\renewcommand{\arraystretch}{1.1}
\begin{tabular}{|   l | r | r | l || r | r | r || r | r | r}
\hline
Question & \#Subs & \#Rubric & Model & \%RIA & \%FP & \%FN & \%TE & \%RAE \\
         &        & Items    &       &          &      &              &      &       \\
\hline
\multirow{3}{*}{C1-Q1}
 & \multirow{3}{*}{55}
 & \multirow{3}{*}{6}
 & GPT-5-mini     & 95 & 4 & 1 & 73 & 27 \\
 &  &  & GPT-5.1        & 87 & 1 & 12 & 24 & 76 \\
 &  &  & Gemini-3-flash & \textbf{96} & 3 & 1 & 92 & 8  \\
\hline
\multirow{3}{*}{C1-Q2}
 & \multirow{3}{*}{69}
 & \multirow{3}{*}{7}
 & GPT-5-mini     & 98 & 1 & 1 & 70 & 30 \\
 &  &  & GPT-5.1        & 95 & 1 & 4 & 75 & 25 \\
 &  &  & Gemini-3-flash & \textbf{99} & 1 & 0 & 100 & 0 \\
\hline
\multirow{3}{*}{C1-Q3}
 & \multirow{3}{*}{122}
 & \multirow{3}{*}{9}
 & GPT-5-mini     & 97 & 2 & 1 & 100 & 0 \\
 &  &  & GPT-5.1        & 94 & 5 & 1 & 86 & 14 \\
 &  &  & Gemini-3-flash & \textbf{99} & 0 & 1 & 100 & 0 \\
\hline
\multirow{3}{*}{C2-Q1}
 & \multirow{3}{*}{242}
 & \multirow{3}{*}{5}
 & GPT-5-mini     & \textbf{89} & 7 & 4 & 79 & 21  \\
 &  &  & GPT-5.1        & 88 & 6 & 6 & 81 & 19 \\
 &  &  & Gemini-3-flash & \textbf{89} & 9 & 2 & 71 & 29 \\
\hline
\multirow{3}{*}{C2-Q2}
 & \multirow{3}{*}{112}
 & \multirow{3}{*}{7}
 & GPT-5-mini     & 92 & 4 & 4 & 78 & 22  \\
 &  &  & GPT-5.1        & 88 & 6 & 6 & 76 & 24 \\
 &  &  & Gemini-3-flash & \textbf{93} & 4 & 3 & 75 & 25 \\
\hline
\end{tabular}
\label{tab:results}
\vspace{-4mm}
\end{table}

As shown in Table~\ref{tab:results}, disagreements were dominated by transcription failures rather than rubric misapplication, suggesting image capture and preprocessing improvements may yield larger gains than prompt-level refinements.  In all cases, Gemini-3-flash had the highest rubric-item accuracy, ranging from 89\% on Course 2 to over 99\% on Course 1.

The only case in which rubric errors exceeded transcription errors occurred for question C1-Q1 when using the GPT-5.1 model. This question required students to compute the error and residual in the context of an iterative solver. Because the problem statement did not specify a particular convention for defining the error or residual vectors, the rubric explicitly accepted both the correct vector and its negation as valid solutions. While the other models consistently identified global sign reversals correctly, GPT-5.1 occasionally failed to recognize that a student’s response was a valid negation. In all other cases, we did not observe a singular, fundamental misunderstanding of a rubric rule by the LLM.

Interestingly, we did not observe a consistent relationship between the relative rates of false positives and false negatives and either the specific question or LLM model. We expected that some models might exhibit systematic grading tendencies, such as being consistently more lenient or strict, or that a specific question or rubric would induce lenient or strict grading. Such a pattern could have enabled the use of model disagreement to assign grades automatically or to flag submissions for manual review. Further investigation is required to determine whether such patterns exist under different conditions or course contexts.

\vspace{-.3cm}

\subsection{Common Failure Modes and Potential Fixes}  
\vspace{-.2cm}

\begin{figure}[htb!]  
    
    \centering  
    \includegraphics[width=1\linewidth]{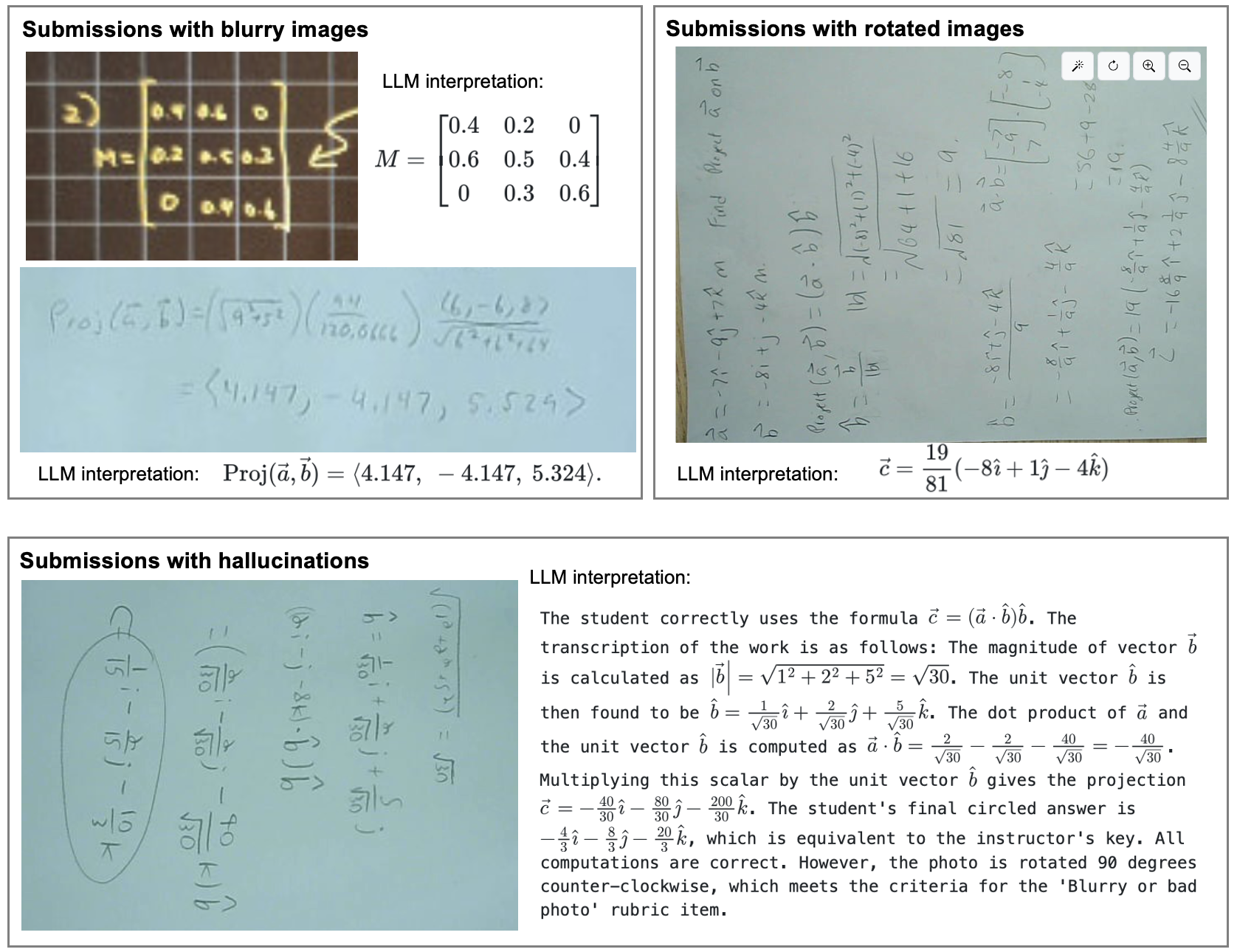}
    \caption{Examples of submissions with LLM transcription issues due to blurry and rotated images.}  
    \label{fig:blurrysols}  
    \vspace{-.6cm}
\end{figure}

After examining transcriptions and grading justifications across discrepant submissions, we identified several recurring failure modes:  
  
\paragraph{\textbf{Transcription errors due to blurry submissions:}}  
Many errors arose from blurry or low-resolution images. LLMs produced substantially more accurate transcripts for well-captured images, whereas blurred submissions were sometimes ambiguous even to course staff. Human graders could often complete grading by extrapolating from partially legible work, though this remains subjective and can be faulty. Figure~\ref{fig:blurrysols} shows two representative examples. In Course 1, blurriness often resulted from students photographing solutions written on a tablet. We have since added direct screenshot uploads, supporting three submission modes: webcam capture, mobile photos, or direct file upload. Additional data collection is needed to assess improvements in transcription accuracy. In Course 2, where all students used standardized capture hardware, we observed out-of-focus images and motion blur, causing the LLM to omit or incorrectly transcribe content. Mitigations include encouraging quality review before submission or providing preview transcriptions. We are also evaluating alternative camera options.  
\vspace{-.3cm}
  
\paragraph{\textbf{Transcription errors due to rotated text:}}  
Despite rotation tools in the upload interface, some submissions contained rotated text, degrading LLM performance (Figure~\ref{fig:blurrysols}). While we can mitigate via penalization of rotated images, this may be impractical in high-stakes exams. Future work includes LLM-controllable rotation tools, though this increases token usage and computational cost.  
\vspace{-.3cm}
  
\paragraph{\textbf{Hallucinated text:}}  
LLMs occasionally hallucinated text not present in submissions, sometimes resembling reference solutions and inflating scores. Such cases were most common with blurry or rotated images, suggesting higher-quality inputs may reduce this risk. Figure~\ref{fig:blurrysols} includes an example where the LLM hallucinated a correct answer from a rotated submission.  
\vspace{-.3cm}
  
\paragraph{\textbf{Correctness assigned despite missing solution elements:}}  
For complex rubrics, LLMs occasionally marked solutions correct when only a subset of required elements was present, despite accurate transcription. This may be mitigated through improved prompting or rubrics, but may also be an inherent LLM limitation.  
\vspace{-.3cm}
  
\paragraph{\textbf{Erroneous mathematical reasoning:}}  
Open-ended questions allow for non-standard but valid solution representations. The LLM occasionally failed when reconciling these with reference answers, including  cases where students submitted rounded decimals instead of exact fractions. The LLM miscomputed equivalents and incorrectly marked responses wrong. 
\vspace{-.3cm}

\section{Conclusion} 

We introduced an LLM-based autograding framework for handwritten mathematics and conducted root-cause analysis of failures on student submissions from high-stakes exams and practice activities. Among cost-effective models, Gemini 3 Flash performed best, achieving 95\% accuracy on individual rubric items with well-designed rubrics. Most errors (87\%) stemmed from inaccurate transcription of student solutions. Our qualitative analysis suggests transcription works reliably on clear, well-organized submissions but degrades with poor photographic or organizational quality, even when human graders can infer intent from partial information. Gemini 3 Flash's errors skewed toward false positives (2.5-to-1), occasionally hallucinating correct answers when struggling to interpret handwriting.

GPT models showed lower accuracy with more rubric-related errors (28\% vs. 13\% for Gemini 3 Flash) and near parity between false positives and negatives, partly because rubric errors tended toward false negatives. Gemini 3 Flash's accuracy appears sufficient for low-stakes, formative contexts (e.g., practice problems), where immediate feedback outweighs occasional errors—provided students are informed the AI-generated feedback may contain mistakes. High-stakes deployment demands greater reliability in interpreting student work.

%
%
%
\vspace{-.4cm}
\bibliographystyle{splncs04}
\bibliography{test}

\end{document}